\documentclass{epl}

\title{Elastic Response of Rough Surfaces in Partial Contact}
\author{G.\ George Batrouni\inst{1} 
\and Alex Hansen\inst{2,3} 
\and Jean Schmittbuhl\inst{4}}
\institute{
\inst{1}Institut Non-Lin\'eaire de Nice, UMR CNRS 6618, 
Universit{\'e} de Nice-Sophia Antipolis, 1361 Route des Lucioles,
F--06560 Valbonne, France\\
\inst{2}Department of Physics,
NTNU, N--7491 Trondheim, Norway\\
\inst{3} NORDITA and Niels Bohr Institute, Blegdamsvej 17, DK--2100
Copenhagen {\O}, Denmark\\
\inst{4}Departement de G{\'e}ologie, UMR CNRS 8538,
Ecole Normale Sup{\'e}rieure,
24, rue Lhomond,
F--75231 Paris C{\'e}dex 05, France
}
\pacs{62.20.Qp}{Tribology and hardness}
\pacs{68.35.Ct}{Interface structure and roughness}
\pacs{91.60.Ba}{Elasticity, fracture and flow}

\begin{document}
\maketitle
\begin{abstract} 
We model numerically the partial normal contact of two elastic rough
surfaces with highly correlated asperities. Facing surfaces are
unmated and described as self-affine with a Hurst exponent $H$. The
numerical algorithm is based on Fourier acceleration and allows
efficient simulation of very large systems. The force, $F$, versus
contact area, $A$, characteristics follows the law $F\sim A^{(1+H)/2}$
in accordance with the suggestion of Roux et al.\ (Europhys.\ Lett.\
{\bf 23}, 277 (1993)).  However finite size corrections are very large
even for $512\times 512$ systems where the effective exponent is still
20\% larger than its asymptotic value.
\vspace{0.3cm}

\noindent

\end{abstract} 
\vskip0.5cm
The mechanical properties of bodies in contact have been studied for a
long time \cite{j85} since they have important applications ranging
from the flow properties of powders to earthquake dynamics
\cite{s90}.  For example, frictional properties are related to the
normal stresses that develop during contact, and they have recently
been shown to be very sensitive to heterogeneities in the normal
stresses \cite{dk96}.  On the fault scale, the dynamical stress field,
which is responsible for earthquakes, is strongly influenced by
heterogeneities due to asperity squeeze \cite{bcc98}.

In this letter we investigate numerically the elastic response of
self-affine rough surfaces that are squeezed together.  The rough
surface, which we take to be oriented in the $(x,y)$ plane, is given
by $z=z(x,y)$.  If $p(z;x,y)$ is the probability density to find the
surface at a height $z$ at $(x,y)$, self affinity is the scaling
property
\begin{equation}
\label{sa}
\lambda^Hp(\lambda^H z; \lambda x,\lambda y)=p(z;x,y)\;,
\end{equation}
where $H$ is the Hurst or roughness exponent \cite{f88,bs97}. There is
strong experimental evidence that surfaces resulting from brittle
fracture are self affine with a Hurst exponent equal to $0.8$
\cite{bs85a,ptbbs87,blp90,mhhr92,sss95,b97}.

When two elastic media are forced into contact along two non-matching
self-affine rough surfaces, two mechanisms conspire to produce a power
law dependence of the applied normal force $F$ on the penetration
depth, $D$: (1) the contact area increases as $D$ is increased, and
(2) the fluctuations in the normal stress field where there is contact
reflect the self affinity of surfaces.  In the much simpler case of a
spherical asperity in contact, Hertz showed about 120 years
ago \cite{ll58} that $F\sim D^{3/2}$, and $A\sim D$.  The Hertz law has
been verified experimentally in the case of a diamond stylus sliding
on a diamond surface\ \cite{bowden}. The power law dependence of the
force on penetration has also been established experimentally for
fracture surfaces in contact \cite{om85}.

In 1993, Roux et al.\ \cite{rsvh93} proposed the following dependence
of $F$ and $A$ on $D$ for self-affine elastic surfaces in contact,
\begin{equation}
\label{fvsdr}
F\sim D^{1+1/H}\;,
\end{equation}
and 
\begin{equation}
\label{avsdr}
A\sim D^{2/H}\;,
\end{equation}
where $H$ is the smallest Hurst exponent of the two surfaces.  The
argument goes as follows.  First, we note that the rough surface with
the smallest Hurst exponent is the roughest and will dominate the
scaling properties of the common interface of the two media in contact
up to a cross-over length scale \cite{pkhrs95}.  Second, we note that
nowhere in the arguments that follow will we need both surfaces to
deform.  Thus, we will make the assumption that the surface with the
largest Hurst exponent (or one of them if they have the same Hurst
exponent) simply is flat and elastic, whereas the other one is rough
and infinitely rigid.  Let us now rescale the spatial coordinates of
the surface, $x\to\lambda x$, $y\to\lambda y$ and $z\to\lambda^H z$.
As a consequence of Eq.\ (\ref{sa}), the surface is statistically
invariant under this operation.  The penetration $D$ needs to be
rescaled $D\to \lambda^H D$ as it ``points'' in the $z$ direction, and
the contact area $A\to \lambda^2 A$ as it lies in the $(x,y)$ plane.
The local deformation, $u$, of the surface must be rescaled as $u\to
\lambda^H u$, since $u$ also ``points'' in the $z$ direction.  The
local deformation, $u$, at $(x,y)$ is related to the normal stress
field $\sigma$ by the expression \cite{ll58}
\begin{equation}
\label{gruf}
u(x,y)=\int\int dxdy G(x-x',y-y') \sigma(x',y')\;,
\end{equation}
where $G\sim1/r$, and $r=|(x-x',y-y')|$.  Thus, we find
that $\sigma\to\lambda^{H-1}\sigma$ under rescaling \cite{hsb00}.  The
total force $F=A\sigma$, and consequently, $F\to\lambda ^{1+H} F$
under rescaling.  Eliminating $\lambda$ between any pair of variables
in the above expressions, gives the power law dependence of the
different variables.  In particular, we find Eqs.\ (\ref{fvsdr}) and
(\ref{avsdr}) and
\begin{equation}
\label{fvsar}
F\sim A^{(1+H)/2}\;.
\end{equation}

Eq.\ (\ref{fvsdr}) was tested experimentally and numerically in Ref.\
\cite{mwhr93}.  The obtained results were consistent with the
suggestions of Roux et al.\ \cite{rsvh93}, but hardly convincing. The
first problem was that the definition of the zero level of the
penetration is {\it not\/} the point of first contact between the two
surfaces, but rather has to be treated as a free parameter in the data
fits.  Another numerical problem encountered was the necessity to
invert dense matrices, with the consequence that only small systems
could be studied (up to $32\times 32$).

\begin{figure}
\centerline{\includegraphics[height=3cm]{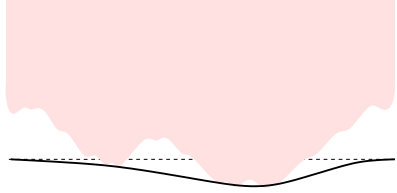}}
\caption{A stiff medium with a rough surface (top) is pushed into an 
elastic, flat medium (bottom).  The dashed, horizontal line represents
the undeformed medium at a given penetration depth, while the full
line represents the actual surface of the medium.  The penetration
depth $D$ is defined with respect to the dashed line.
\label{fig1}}
\end{figure}

In the present letter, we have developed a new and much more efficient
algorithm that allows us to simulate easily systems up to $512\times
512$. We are, therefore, able to present convincing tests of the
scaling relation Eq.\ (\ref{fvsar}).  Whereas the numerics of Ref.\
\cite{mwhr93} were based on six samples of size $32\times 32$, we base
our data on $300$ samples of size $512\times 512$.  One surprising
result in the present study is how strong the finite-size corrections
to the results are. Size is estimated in units of the lower cut-off
scale for the self-affine invariance of the rough surfaces. 

We now describe the model and its numerical solution.  We represent
both surfaces on two-dimensional $L\times L$ lattices with
deformations taking place only in the third transverse direction. The
discretization size is adjusted to the unit size that is the lower
cut-off length of the self-affine scaling. As the hard rough surface
is pushed into the elastic one, the forces and deformations are
described by the discrete version of Eq.\ (\ref{M1}),
\begin{equation}
\label{M1}
u_i=\sum_j G_{i,j} f_j\;,
\end{equation}
with the Green function given by
\begin{equation}
\label{M2}
G_{i,j} = {{2(1-s)(1+s)} \over 
{2\pi e\ |{\vec i}-{\vec j}|}}\;.
\end{equation}
$u_i$ is the deformation of the elastic body at site $i$ and $f_i$
the force acting at that point. In the Green function, $s$ is the
Poisson ratio, $e$ the elastic constant, and $|{\vec i}-{\vec j}|$ the
distance between sites $i$ and $j$. The indices $i$ and $j$ run over
all $L^2$ sites.

To define the problem completely, the boundary conditions need to be
specified. Clearly, in the regions where the rough surface is in
contact with the elastic one, the {\it deformation\/}, $u_i$, is
specified (it conforms to the shape of the rough surface) and one
solves for the force. On the other hand, in the regions where contact
has not been established, the elastic surface deforms in response to
the influences from the contact regions. In this case equilibrium is
established when the net forces acting on the surface
vanish. Therefore, in the no-contact region, the force is specified,
it vanishes, and one solves for the deformation.

We can build these boundary conditions into the equations to
facilitate solving them. To do this, we define the {\it diagonal\/}
$L^2\times L^2$ matrix, ${\bf K}$, with elements equal to $1$ on
contact sites and $0$ on free (no-contact) sites. Clearly the vector
${\bf K}{\vec u}$ is zero everywhere there is no contact. At contact
points, ${\bf K}{\vec u}$ is equal to the imposed deformation given by
the shape of the rough surface.

Eq.\ (\ref{M1}) can be rewritten as,
\begin{equation}
\label{M3}
{\bf G}({\bf I}-{\bf K}){\vec f}+{\bf G}{\bf K}{\vec f}=({\bf
I}-{\bf K}){\vec u}+{\bf K}{\vec u}\;,
\end{equation}
where we use matrix-vector notation, and ${\bf I}$ is the
identity. This form is convenient because as mentioned above, ${\bf
K}{\vec u}$ is a known quantity (boundary condition). In addition, the
vector $({\bf I}-{\bf K}){\vec f}$ is always zero because the force
${\vec f}$ is nonzero only at contact points. Putting the unknowns on
the left hand side and the boundary conditions on the right hand side
of Eq.\ (\ref{M3}) we obtain,
\begin{equation}
\label{M4}
{\bf G}{\bf K}{\vec f}-({\bf I}-{\bf K}){\vec u}={\bf K}{\vec u}\;.
\end{equation}
Now define the vector ${\vec x}$ representing all the unknown
quantities. Clearly,
\begin{equation}
\label{M5}
{\vec x} = {\bf K}{\vec f} + ({\bf I}-{\bf K}){\vec u}\;.
\end{equation}
With this definition, and noting that $ {\bf K}({\bf I}-{\bf K})=({\bf
I}-{\bf K}){\bf K}=0$, ${\bf K}^2={\bf K}$, and $({\bf I}-{\bf
K})^2=({\bf I}-{\bf K})$ we can write Eq.\ (\ref{M4}) as
\begin{equation}
\label{M6}
{\bf G}{\bf K}{\vec x}-({\bf I}-{\bf K}){\vec x}={\bf K}{\vec u}\;,
\end{equation}
and finally
\begin{equation}
\label{M7}
\bigl ({\bf I}-({\bf I}-{\bf G}){\bf K}\bigr ){\vec x}={\bf K}{\vec
u}\;.
\end{equation}

Eq.\ (\ref{M7}) is of the familiar form, ${\bf A}{\vec x}={\vec b}$
which can be solved using, for example, the Conjugate Gradient (CG)
method \cite{ptvf92,bh88}. The difficulty is that the Green function
${\bf G}$ is represented by a {\it full\/} $L^2\times L^2$
matrix. This will lead to the number of operations per iteration
scaling as $L^4$ which is prohibitive. We overcome this difficulty by
doing the matrix multiplications involving ${\bf G}$ in Fourier space
by using FFTs. This leads to a stable and extremely efficient
algorithm, scaling like $L^2\ln(L)$, with which we were able to study
systems with size up to $L=512$.

One more technical detail remains. The matrix ${\bf K}$, which
indicates the contact points needs to be determined. The problem is
that as we push the rough surface into the elastic one, the latter
deforms. Therefore, the contact area is not equal to the area obtained
by simply taking a cut through the rough surface. We obtain the
correct contact area as follows. Our initial assumption is that the
contact area is equal to the area obtained from a simple cut of the
rough terrain, see Fig.\ \ref{fig1}. This determines the initial ${\bf
K}$ which is then used in Eq.\ (\ref{M7}). The solution thus obtained
gives the forces where there is contact and the deformations where
there is none. Some of the forces thus obtained are negative since the
elastic surface is trying to pull away from the rough surface. We
therefore modify ${\bf K}$ by zeroing the elements corresponding
to sites where the force is negative, and we solve again. We repeat
this process until there are no sites with negative forces. This
algorithm always converges giving the correct contact area and forces.

\begin{figure}
\centerline{\includegraphics[height=8cm,width=6cm,angle=270]{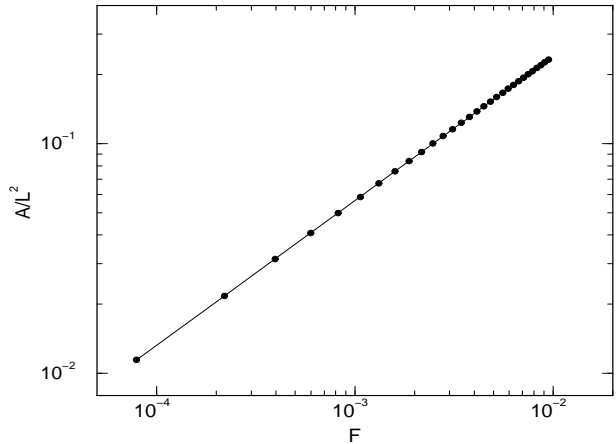}}
\caption{Contact area versus force for a hard sphere with radius $R=6000$
squeezed into an elastic flat medium with elastic constant $e=100$
({\it i.e. a Hertz contact}).  The calculation is done on a $512\times
512$ lattice.  The least-squares fit, assuming a power law gives
$A\sim F^{0.63}$.
\label{fig2}}
\end{figure}

To verify the algorithm and program we first tested it with a Hertz
contact. Figure\ \ref{fig2} shows the force-contact area
characteristics of a hard sphere with radius $R=6000$ which is pushed
into an elastic, flat medium with elastic constant $e=100$.  The
calculation was done on a $512\times 512$ lattice.  The exact Hertz
solution gives $A\sim F^{2/3}$, while a least-squares fit on the data
gives $A\sim F^{0.63}$.  The exponent given by our model, $0.63$, is
in excellent agreement with the exact value, the difference being due
to finite size effects. We have verified that for smaller systems and
smaller asperities, the exponent moves farther away from the $2/3$
value.

Having established that the algorithm works correctly, we then
simulated various aspects of squeezing a stiff self-affine surface
into an initially flat elastic one. We measured the total force, $F$,
as a function of the total contact area, $A$. In addition, for all
situations simulated, we measured the force at each contact point thus 
obtaining a very detailed picture of the squeezing process.

\begin{figure}
\hspace*{0.5cm}\includegraphics[height=9cm,height=6cm,angle=270]{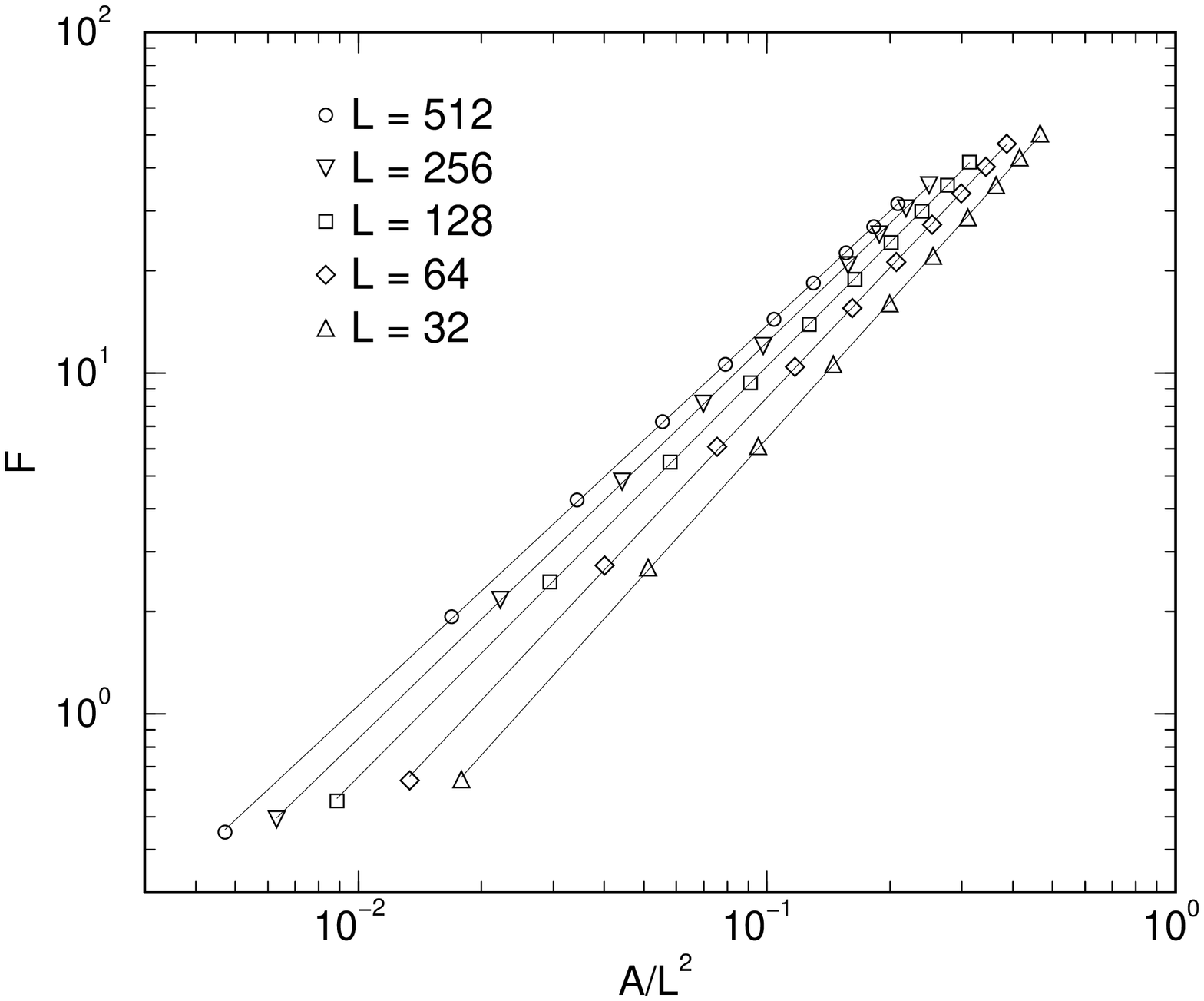}
\hspace*{1cm}\includegraphics[height=9cm,height=6cm,angle=270]{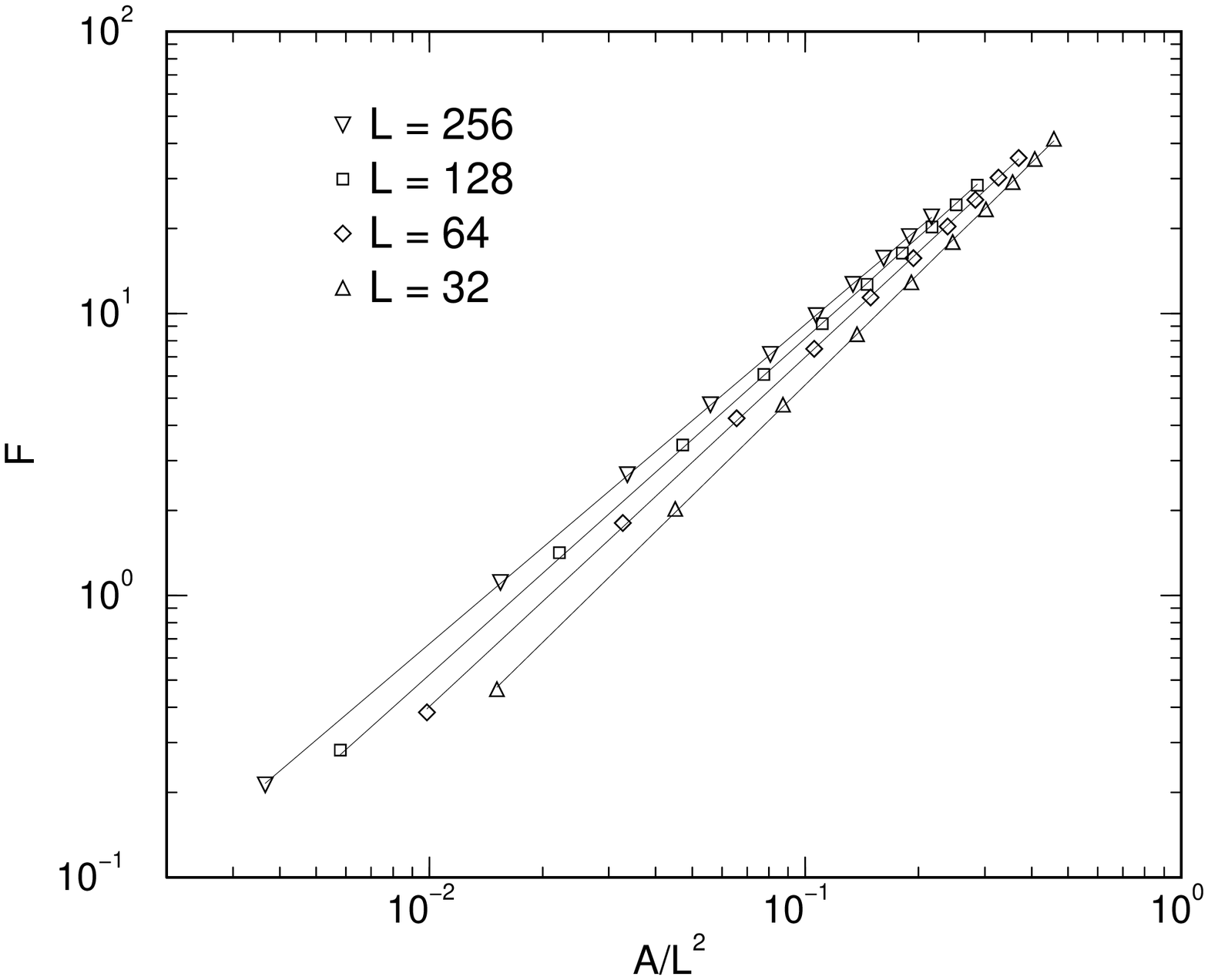}
\caption{{\bf left:} Force, $F$, versus contact area, $A$, for 300
independent samples of hard self-affine surfaces with $H=0.8$ pushed
into an initially flat medium with elastic constant $e=100$. System
sizes from bottom to top: $32$, $64$, $128$, $256$ and $512$. {\bf
right:} Same but with $H=0.6$ and sizes up to $256$.}
\label{fig3}
\label{fig4}
\end{figure}

We show in Fig.\ \ref{fig3} the force, $F$, versus contact area, $A$,
for different system sizes and Hurst exponent $H=0.6, 0.8$. The data
are averaged over $300$ samples for each size and $H$.  The elastic
constant for the elastic, flat medium is $e=100$.  Even though the
force versus contact area curves produce very high-quality fits to
power laws, $F\sim A^\alpha$, the resulting exponents, $\alpha$,
depend strongly on the system size, $L$, even for systems as large as
$512\times 512$.  For $H=0.8$, we find from Fig.\ \ref{fig3},
$\alpha(L=32) =1.33$, $\alpha(64)=1.27$, $\alpha(128)=1.21$,
$\alpha(256)=1.16$, and $\alpha(512)=1.12$.  Correspondingly, from
Fig.\ \ref{fig4}, we find for $H=0.6$, $\alpha(L=32)=1.31$,
$\alpha(64)=1.24$, $\alpha(128)=1.19$, and $\alpha(256)=1.13$.

\begin{figure}
\centerline{\includegraphics[height=8cm,width=6cm,angle=270]{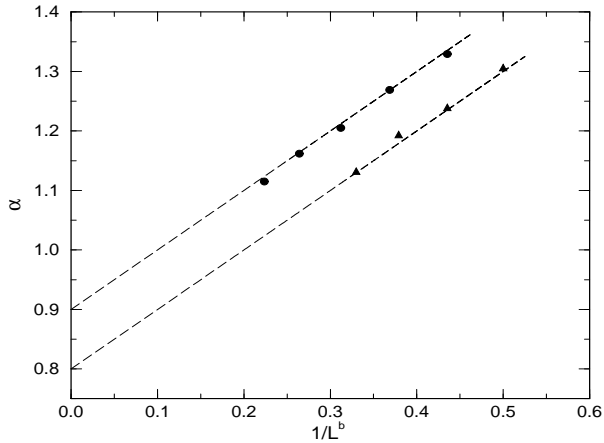}}
\caption{Effective exponent $\alpha$ versus $1/L^b$, where $L$ is
the system size. Circles are for $H=0.8$ and give $b=0.24$, triangles
are $H=0.6$ and give $b=0.2$. Curves are given by $\alpha=(1+H)/2 +
L^{-b}$.
\label{fig5}}
\end{figure}

In figure \ref{fig5} we show these exponents versus $1/L^b$ for
$H=0.8$ (circles) and $H=0.6$ (triangles). It is clear from this
figure that the finite size effects are large. It is not easy to
extrapolate to $L\to \infty$ using $\alpha(L) = \alpha_{\infty} + a
L^{-b}$ because the power law has a small exponent, $b$, and requires
much larger sizes to give reasonable results to this three parameter
fit. Instead, we will verify the result of reference\ \cite{rsvh93} by
fitting to a one parameter function, $\alpha(L) = (1+H)/2+L^{-b}$. The
results are shown as solid lines in Fig.\ \ref{fig5}. The high quality
of the fits strongly supports the result $\alpha_{\infty}=(1+H)/2$. We
found $b=0.24$ for $H=0.8$ and $b=0.2$ for $H=0.6$. It is not clear if
the difference between the two exponents is meaningful.

It is worth noting that although the case of the Hertz contact did
display some finite size effects, they were very small especially when
compared with the large effects observed with self affine
surfaces. This can be attributed to the long distance correlations in
self affine surfaces.

Thanks to a fast algorithm based on Fourier acceleration we were able
to describe precisely the evolution of the contact between two elastic
rough surfaces from the first contact of one asperity to the maximum
contact (see below). Asperity roughness is characterized by a
self-affine (Hurst) exponent in a range compatible with fractured
surfaces. Our algorithm easily allows extension to other roughness
exponents. We demonstrated that the theoretical prediction from Roux
{\em et al} \cite{rsvh93} is valid but very strong finite size effects
exist. For instance, the effective relationship between normal force
and contact area is
\begin{equation}
F\propto A^{(1+H)/2 + L^{-b}}
\end{equation}
were $b\approx 0.2$. This finite size effect results in a large
sensitivity to the cut-off scale of the self-affine scaling.

Finally, we remark that when we push a hard rough body into the
elastic initially flat one, the maximum contact area, $A/L^2$,
obtained is never unity! In Fig.\ \ref{fig3} we show the total force,
$F$, versus the contact area, $A/L^2$, {\it up to the maximum
attainable contact area}. It is clear from the figure that it is not
unity: For the $L=512$ system it is of the order of $20\%$. If we
allow plastic yield, full contact would be attainable. This is
straightforward to do in our model.

\acknowledgments

We thank F.\ A.\ Oliveira and H.\ Nazareno and the ICCMP of the
Universidade de Bras{\'\i}lia for friendly support and hospitality
during the initial phases of this project.  This work was partially
funded by the CNRS PICS contract $\#753$, the Norwegian research
council (NFR) and NORDITA. A.\ H.\ also thanks the Niels Bohr
Institute for hospitality and support.

\end{document}